\documentclass[twocolumn,prl,aps,showpacs,amsmath,amssymb]{revtex4}

\usepackage{graphicx}
\usepackage{dcolumn}
\usepackage{bm}

\renewcommand{\vec}[1]{{\bf#1}}

\begin{document}

\title{Extrinsic and intrinsic ratchet response of a quantum dissipative spin-orbit medium}

\author{Sergey Smirnov,$^1$ Dario Bercioux,$^2$ Milena Grifoni,$^1$ and Klaus Richter$^1$}
\affiliation{$^1$Institut f\"ur Theoretische Physik, Universit\"at Regensburg, D-93040 Regensburg, Germany\\
  $^2$Freiburg Institute for Advanced Studies (FRIAS) and Physikalisches Institut, Universit\"at Freiburg, D-79104
  Freiburg, Germany}

\date{\today}

\begin{abstract}
Traditionally the charge ratchet effect is considered as a consequence of the extrinsic spatial asymmetry engineered by
external asymmetric periodic potentials. Here we demonstrate that electrically and magnetically driven dissipative systems
with spin-orbit interactions represent an exception from this standard idea. The charge and spin ratchet currents appear
just due to the coexistence of quantum dissipation with the intrinsic spatial asymmetry of the spin-orbit coupling. The
extrinsic spatial asymmetry is inessential.
\end{abstract}


\maketitle

 A system of charged particles driven by a time-dependent external force may exhibit a net charge current even if the
force is periodic and unbiased. This so-called charge ratchet effect \cite{Astumian,Reimann,Juelicher,Linke,Majer,Haenggi}
is used e.g. in nano-generators of direct currents. If transport involves the spin degree of freedom, the concept of a
spin ratchet \cite{Scheid,Scheid_1,Smirnov,Smirnov_1} emerges as a natural analog of the charge ratchet notion.

For systems with spin-orbit interactions the {\it spin} ratchet effect may have been expected because it could be rooted
in an asymmetric excitation of spin dynamics by the orbital dynamics induced by an electric field. Such an expectation is
based on the intrinsic spatial asymmetry inherent to systems with spin-orbit interactions. For example the Rashba
\cite{Rashba} and Dresselhaus \cite{Dresselhaus} spin-orbit Hamiltonians for semiconductor heterostructures are obviously
not invariant with respect to the real space inversion. For electrically driven coherent and dissipative systems with
Rashba spin-orbit interaction (RSOI) the spin ratchet mechanism has indeed been confirmed \cite{Scheid,Smirnov,Smirnov_1}.
Even for symmetric periodic potentials the spin ratchet effect exists \cite{Scheid} just due to the intrinsic spatial
asymmetry of RSOI. However, the {\it charge} ratchet effect is absent in both the coherent and dissipative cases for
symmetric periodic potentials. This could deepen the impression that without the extrinsic asymmetry a system will never
respond to external fields via the charge ratchet mechanism and systems with spin-orbit interactions like all other
systems obey this habitual rule. The present work reveals that this is a delusion and in reality systems with spin-orbit
interactions provide a unique opportunity to answer the fundamental questions related to the role of symmetries in ratchet
phenomena in general.

In this Letter we show that the extrinsic asymmetry, usually required as a key property of particle ratchets, is not
necessary as the intrinsic Rashba asymmetry alone is sufficient if a dissipative system is driven by both electric and
magnetic fields. Specifically, it is found that the charge and spin ratchet effects in this case exist for symmetric
periodic potentials and stem just from the simultaneous presence of dissipation and the real space asymmetry of the Rashba
electrons. We also find that at low temperatures the ratchet charge current in the system is unusual. Its queerness
consists in the fact that this current, in contrast to early predictions for systems without spin-orbit interactions
\cite{Grifoni,Goychuk}, appears even when only one energy band provides electrons for transport and no harmonic mixing is
present in the driving fields. This charge current is of pure spin-orbit nature and, as a result, it disappears when the
spin-orbit coupling strength vanishes. Therefore such spin-orbit charge currents can be controlled by the same gate
voltage which controls the strength of the spin-orbit coupling in the system. This could be very attractive from an
experimental point of view since measurements of charge currents are experimentally better controlled than measurements of
spin currents.

\begin{figure}
\includegraphics[width=5.8 cm]{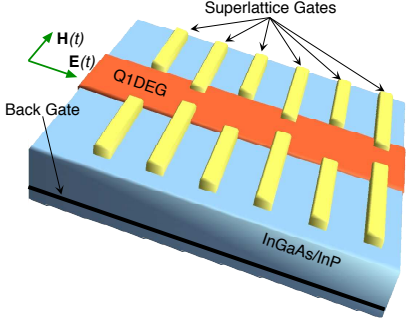}
\caption{\label{figure_1} (Color online) A 2DEG with RSOI of strength $\alpha=9.94\times 10^{-12}$ eV$\cdot$m is
obtained by a gate voltage applied to an InGaAs/InP heterostructure using the "Back Gate". The electron effective mass is
$m=0.037m_0$ with $m_0$ being the free electron mass and the effective gyroscopic factor is $g^*=-15$. A parabolic
confinement of strength $\hbar\omega_0=0.225$ meV forms in the 2DEG a quasi-one-dimensional electron gas (Q1DEG). The
superlattice with period $L=0.25\,\mu m$ is shaped by the "Superlattice Gates" which create a periodic potential whose
strength varies across the Q1DEG producing a coupling between the transverse and longitudinal electron orbital degrees of
freedom. The "Superlattice Gates" can be electrically reprogrammed and thus one can easily switch between symmetric and
asymmetric periodic potentials. The system is driven by a longitudinal electric field $\vec{E}(t)$ and by a transverse
magnetic field $\vec{H}(t)$.}
\end{figure}

An archetype of the device under investigation is shown in Fig.~\ref{figure_1}. In this system non-interacting electrons
are confined in a quasi-one-dimensional (quasi-1D) periodic structure obtained by appropriately placed gates applied to a
two-dimensional electron gas (2DEG) with RSOI. The system interacts with an external environment (or bath): the
longitudinal orbital degree of freedom of each electron is coupled to orbital degrees of freedom of the external
environment. This coupling is the source of dissipation in the system. The electrons are driven by longitudinal electric
and transverse in-plane magnetic homogeneous fields which are time-periodic functions with zero mean value.

To perform a quantitative analysis of the ratchet effects we model the system by the Hamiltonian $\hat{H}(t)=\hat{H}_0+
\hat{H}_\mathrm{D}(t)+\hat{H}_\mathrm{B}$, where $\hat{H}_\mathrm{D}(t)\equiv -eE(t)\hat{x}-
g\mu_\mathrm{B}H(t)\hat{\sigma}_z$ is the driving term, $\hat{H}_\mathrm{B}$ is the bath term of the Caldeira-Leggett model
\cite{Caldeira,Weiss} taking into account the orbital coupling between the electron longitudinal degree of freedom,
$\hat{x}$, and orbital degrees of freedom of the bath. All properties of the bath are encapsulated in its spectral density
$J(\omega)$. Finally, $\hat{H}_0$ is the Hamiltonian of the isolated system:
\begin{equation}
\hat{H}_0\equiv\frac{\hbar^2\hat{\vec{k}}^2}{2m}-
\frac{\hbar^2k_{\mathrm{so}}}{m}\bigl(\hat{\sigma}_x\hat{k}_z-\hat{\sigma}_z\hat{k}_x\bigl)+V(\hat{x},\hat{z}),
\label{HIS}
\end{equation}
where $V(x,z)\equiv m\omega_0^2z^2/2+U(x)(1+\gamma z^2/L^2)$. In this model it is assumed that the 2DEG is in the $x-z$
plane and the quasi-1D structure is formed along the $x$-axis using a harmonic confinement of strength $\omega_0$ along the
$z$-axis. The electron spin $g$-factor is denoted as $g$ and $\mu_\mathrm{B}$ is the Bohr magneton. The super-lattice
period is $L$, $U(x+L)=U(x)$. The parameters $k_\mathrm{so}\equiv\alpha m/\hbar^2$ and $\gamma$ characterize the strength
of the spin-orbit and orbit-orbit couplings, respectively. We consider the additional possibility of coupling between the
longitudinal and transverse orbital degrees of freedom of the electrons since it is responsible for the existence of the
ratchet transport in the system under appropriate combinations of the driving fields.

The electric driving is given by the vector $\vec{E}(t)=(E(t),0,0)$ while the magnetic driving is $\vec{H}(t)=(0,0,H(t))$.
We consider the time dependence $eE(t)\equiv F\cos(\Omega(t-t_0))$, $H(t)\equiv H\cos(\Omega(t-t_0))$. The vector potential
is chosen using the Landau gauge $\vec{A}(t)=(-H(t)y,0,0)$. Since $y=0$ in the 2DEG, the vector potential is not
explicitly present in the model.

To study the ratchet effects at low temperatures when only the lowest energy band of the super-lattice is populated with
electrons we calculate the charge and spin currents averaged over one driving period. These currents in the long time
limit provide the stationary ratchet response of the system. The common eigenstates of $\hat{x}$ and $\hat{\sigma}_z$
represent a convenient basis to obtain this response. Because of the discrete eigenvalue structure of $\hat{x}$ (see
below) the basis is called the $\sigma$-discrete variable representation ($\sigma$-DVR) basis. The eigenstates are denoted
as $|m,j,\sigma\rangle$, where $m=0,\pm 1,\pm 2,\ldots$, and $j$ and $\sigma$ are the transverse mode and spin quantum
numbers, respectively \cite{Smirnov,Smirnov_1}. In the $\sigma$-DVR basis the averaged charge and spin currents have a
simple form:
\begin{equation}
\begin{split}
&J_\mathrm{C}=-e\underset{t\to\infty}{\mathrm{lim}}\sum_{m,j,\sigma} x_{m,j}\frac{d}{dt}P_{j,\sigma}^m(t),\\
&J_\mathrm{S}=\underset{t\to\infty}{\mathrm{lim}}\sum_{m,j,\sigma}\sigma x_{m,j}\frac{d}{dt}P_{j,\sigma}^m(t).
\end{split}
\label{CCSC}
\end{equation}
In Eq. (\ref{CCSC}) $P_{j,\sigma}^m(t)$ is the averaged population at time $t$ of the $\sigma$-DVR state
$|m,j,\sigma\rangle$, the quantities $x_{m,j}=mL+d_j$ ($-L/2<d_j\leqslant L/2$) and $\sigma$ are eigenvalues of $\hat{x}$
and $\hat{\sigma}_z$ corresponding to their common eigenstate $|m,j,\sigma\rangle$. Additionally, the $\sigma$-DVR basis
allows the path integral formalism to handle the magnetic driving on an equal footing with the standard electric driving
since in this basis the whole driving Hamiltonian, $\hat{H}_\mathrm{D}(t)$, is diagonal.

In the long time limit the populations $P_{j,\sigma}^m(t)$ come from a master equation \cite{Smirnov,Weiss} which is in this
case Markovian.

Before starting a rigorous exploration one can already anticipate that the magnetic field driving brings a whiff of fresh
physics because the spin dynamics can be controlled directly and not only through the spin-orbit interaction mediating
between the electric field and electron spins.

An analytical treatment of this rather complicated problem is possible when the dynamics of $P_{j,\sigma}^m(t)$ is treated
within the first two transverse modes, i.e., $j=0,1$.

For a detailed study we derive the charge and spin currents assuming weak coupling between neighboring $\sigma$-DVR
states. We obtain:
\begin{equation}
\begin{split}
&J_\mathrm{C}\!\equiv\!
\frac{2eL}{I}\bigl|\Delta_{\uparrow\downarrow}^{01}\bigl|^2\bigl|\Delta_{\downarrow\uparrow}^{10}\bigl|^2
\bigl(I_{\uparrow\downarrow}^{01,\mathrm{b}}I_{\downarrow\uparrow}^{10,\mathrm{b}}-
I_{\uparrow\downarrow}^{01,\mathrm{f}}I_{\downarrow\uparrow}^{10,\mathrm{f}}\bigl),\\
&J_\mathrm{S}\!\equiv\!
\frac{2L}{I}\bigl(\bigl|\Delta_{\uparrow\downarrow}^{01}\bigl|^4I_{\uparrow\downarrow}^{01,\mathrm{f}}
I_{\downarrow\uparrow}^{10,\mathrm{b}}-\bigl|\Delta_{\downarrow\uparrow}^{10}\bigl|^4I_{\uparrow\downarrow}^{01,\mathrm{b}}
I_{\downarrow\uparrow}^{10,\mathrm{f}}\bigl),
\end{split}
\label{CCSCF}
\end{equation}
where $\Delta_{\sigma'\sigma}^{j'j}\equiv\,\langle m+1,j',\sigma'|\hat{H}_0|m,j,\sigma\rangle$ are the hopping
matrix elements of the Hamiltonian of the isolated system, Eq. (\ref{HIS}),
$I\!\equiv\!\bigl|\Delta_{\uparrow\downarrow}^{01}\bigl|^2\bigl(I_{\uparrow\downarrow}^{01,f}\!\!+
\!I_{\downarrow\uparrow}^{10,b}\bigl)+\bigl|\Delta_{\downarrow\uparrow}^{10}\bigl|^2\bigl(I_{\uparrow\downarrow}^{01,b}\!+
\!I_{\downarrow\uparrow}^{10,f}\bigl)$, and $\uparrow,\downarrow$ stand for $\sigma=1,-1$, respectively. The effects of both
the driving fields and quantum dissipation are in the integrals
\begin{equation}
\begin{split}
I_{\sigma'\sigma}^{j'j,\mathrm{\binom{f}{b}}}
&\equiv\frac{1}{\hbar^2}\int_{-\infty}^{\infty}d\tau\mathrm{e}^{-\frac{L^2}{\hbar}
Q(\tau;J(\omega),T)+\mathrm{i}\frac{\tau}{\hbar}(\varepsilon_\sigma^j-\varepsilon_{\sigma'}^{j'})}\times\\
&\times J_0\biggl[\frac{\mp 2FL+2g\mu_\mathrm{B}H(\sigma-\sigma')}{\hbar\Omega}
\sin\biggl(\frac{\Omega\tau}{2}\biggl)\biggl],
\end{split}
\label{DDI}
\end{equation}
where $Q[\tau;J(\omega),T]$ is the twice integrated bath correlation function \cite{Weiss} whose dependence on $\tau$ is
fixed by the bath spectral density $J(\omega)$ and temperature $T$,
$\varepsilon_\sigma^j\equiv\langle m,j,\sigma|\hat{H}_0|m,j,\sigma\rangle$ are the on-site energies of the isolated system,
and $J_0(x)$ is the Bessel function of zero order.

Remarkably, Eq. (\ref{CCSCF}) tells us that at low temperatures the ratchet charge and spin transport in the system exists
just because of spin flip processes. Whereas it looks natural for the spin current, it is a quite unexpected and important
result for the charge current. This current emerges because the magnetic driving changes the charge dynamics. In this case
the spin-orbit interaction plays a role inverse to the one which it plays for the electric driving: the magnetic field
exciting spin dynamics induces orbital dynamics through the spin-orbit interaction. The corresponding charge flow,
originating just due to the spin-orbit interaction, is finite even when only one energy band contributes to transport.

The situation, however, is highly non-trivial and the final conclusions about the existence of the ratchet charge and spin
flows cannot be based only on the presence of spin-orbit interactions. There are also external time-dependent fields
driving the system and internal quantum dissipative processes. The mutual driving-dissipation effect is incorporated in
the integrals, Eq. (\ref{DDI}). Therefore, a further analysis is required: one should additionally take into consideration
the properties of the integrals from Eq. (\ref{DDI}) and the properties of the static periodic potential with respect to
the spatial inversion symmetry.

There are twelve different cases, shown in Table \ref{ECSRE}, to check whether the charge and spin ratchet effects can
take place in the corresponding physical situations. Only those four of them which are given by the row with $F\neq 0$,
$H=0$ have been studied up to now and discussed in Refs. \cite{Smirnov,Smirnov_1}. The other eight possibilities have not
been investigated so far.

\begin{table}[h!b!p!]
\caption{Existence of the charge and spin ratchet effects}
\begin{tabular}{|>{\tiny}m{0.7cm}|>{\tiny}m{1.55cm}|>{\tiny}m{1.55cm}|>{\tiny}m{1.55cm}|>{\tiny}m{1.55cm}|}
\hline
& \multicolumn{2}{>{\tiny}c|}{$\gamma=0$} & \multicolumn{2}{>{\tiny}c|}{$\gamma\neq 0$}\\
\cline{2-5}\raisebox{1.5ex}[0cm][0cm]{}
& $U(x)\neq U(-x)$ & $U(x)=U(-x)$ & $U(x)\neq U(-x)$ & $U(x)=U(-x)$\\
\hline
$F\neq 0$ $H=0$ & \begin{center}{$J_\mathrm{C}=0$ $J_\mathrm{S}=0$}\end{center} &
\begin{center}{$J_\mathrm{C}=0$ $J_\mathrm{S}=0$}\end{center} &
\begin{center}{$J_\mathrm{C}=0$ $J_\mathrm{S}\neq 0$}\end{center} &
\begin{center}{$J_\mathrm{C}=0$ $J_\mathrm{S}=0$}\end{center} \\
\hline
$F=0$ $H\neq 0$ & \begin{center}{$J_\mathrm{C}=0$ $J_\mathrm{S}=0$}\end{center} &
\begin{center}{$J_\mathrm{C}=0$ $J_\mathrm{S}=0$}\end{center} &
\begin{center}{$J_\mathrm{C}=0$ $J_\mathrm{S}\neq 0$}\end{center} &
\begin{center}{$J_\mathrm{C}=0$ $J_\mathrm{S}=0$}\end{center} \\
\hline
$F\neq 0$ $H\neq 0$ & \begin{center}{$J_\mathrm{C}\neq 0$ $J_\mathrm{S}\neq 0$}\end{center} &
\begin{center}{$J_\mathrm{C}\neq 0$ $J_\mathrm{S}\neq 0$}\end{center} &
\begin{center}{$J_\mathrm{C}\neq 0$ $J_\mathrm{S}\neq 0$}\end{center} &
\begin{center}{$J_\mathrm{C}\neq 0$ $J_\mathrm{S}\neq 0$}\end{center}\\
\hline
\end{tabular}
\label{ECSRE}
\end{table}
The results presented in Table \ref{ECSRE} are easily obtained from Eq. (\ref{CCSCF}) if one takes into account that for
$\gamma=0$ or $U(x)=U(-x)$ the equality $|\Delta_{\uparrow\downarrow}^{01}\bigl|=|\Delta_{\downarrow\uparrow}^{10}\bigl|$ is
valid \cite{Smirnov,Smirnov_1}, and for $F=0$ or $H=0$ one makes use of the equality
$I_{\sigma'\sigma}^{j'j,\mathrm{f}}=I_{\sigma',\sigma}^{j'j,\mathrm{b}}$ which follows from Eq. (\ref{DDI}).

\begin{figure*}
\includegraphics[width=18.0 cm]{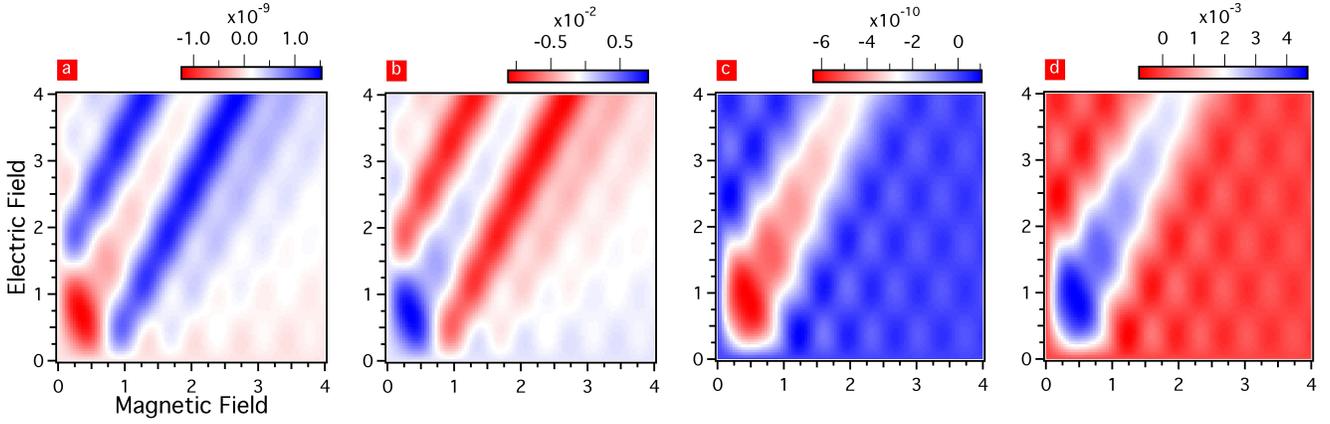}
\caption{\label{figure_2} (Color online) The charge and spin ratchet currents as functions of the amplitudes of the
  electric and magnetic fields. a,b, Spin current for the symmetric and asymmetric cases, respectively. c,d, Charge
  current for the symmetric and asymmetric cases, respectively. The amplitudes of the electric, $FL$, and magnetic,
  $g\mu_\mathrm{B}H$, fields are in units of $\hbar\omega_0$. The currents are in units of $L\omega_0$. The orbit-orbit
  coupling is absent, $\gamma=0$, but the spin current is finite in the symmetric case when both of the fields are
  present. The charge current is excited when both the electric and magnetic fields simultaneously drive the system. In
  the intrinsic ratchet response (a and c) the magnitude of the charge and spin currents is strongly suppressed by the
  symmetry of the periodic potential while in the extrinsic ratchet response (b and d) the charge and spin currents are
  enhanced by the spatial asymmetry of the system.}
\end{figure*}
The principal feature of the physics taking place when $F\neq 0$ and $H\neq 0$ is that the existence of the ratchet
effects is {\it not} dictated only by properties of the isolated system as in Refs. \cite{Smirnov,Smirnov_1}. The physical
picture is now more intricate. In the charge and spin currents one cannot find clear traces of either driving and
dissipation or the isolated system. The two imprints are not separable and the charge and spin ratchet mechanisms are
determined by the whole system-plus-bath complex.

\begin{figure}
\includegraphics[width=6.39 cm]{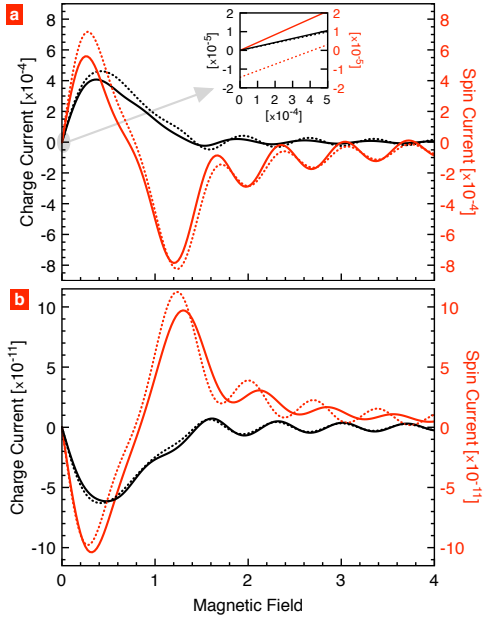}
\caption{\label{figure_3} (Color online) The charge and spin ratchet currents as functions of the magnetic field
  amplitude. The magnetic amplitude, $g\mu_\mathrm{B}H$, is in units of $\hbar\omega_0$. The electric amplitude is fixed,
  $FL=\hbar\omega_0$. The solid curves correspond to $\gamma=0$. The dotted curves correspond to $\gamma=0.1$. a,
  Asymmetric case. The inset shows a vicinity of the point $H=0$ at which the pure spin ratchet response takes place for
  $\gamma\neq 0$. b, Symmetric case.}
\end{figure}

The above theoretical predictions have been confirmed numerically. Figure \ref{figure_2} shows the situation with
$\gamma=0$. The superlattice is modeled by the potential $U(x)=V_0+\sum_{l=1}^2 V_l\cos(2\pi lx/L-\phi_l)$ with
$V_0=-V_1=2.6\hbar\omega_0$, $V_2=1.9\hbar\omega_0$, $\phi_1=1.9$, $\phi_2=0$ for the asymmetric case while for the
symmetric one $V_0=-V_1=2.6\hbar\omega_0$, $V_2=0$, $\phi_1=\phi_2=0$. The period is $L=2.5\sqrt{\hbar/m\omega_0}$ which
gives $k_\mathrm{so}L\approx 0.368\pi$. The driving frequency of the electric and magnetic fields is
$\Omega=\sqrt{3}\omega_0/4$. The bath is Ohmic with the exponential cut-off at $\omega_c=10\omega_0$:
$J(\omega)=\eta\omega\exp(-\omega/\omega_c)$. The viscosity coefficient is $\eta=0.1$ and the temperature is
$k_\mathrm{B}T=0.5\hbar\omega_0$. As theoretically expected the ratchet effects exist even when the periodic potential is
symmetric, Figs.~\ref{figure_2}a and \ref{figure_2}c. However, the currents of these intrinsic ratchet effects are much
smaller than the corresponding currents of the extrinsic ones, Fig.~\ref{figure_2}b and \ref{figure_2}d. What is
surprising in the case when both of the driving fields are present is that the orbit-orbit coupling has a weak effect on
the ratchet spin current as it is demonstrated in Fig.~\ref{figure_3}. At the same time when $H=0$ the orbit-orbit
coupling is responsible for the existence of the pure spin ratchet effect (see Ref. \cite{Smirnov}) as one can see in the
inset of Fig.~\ref{figure_3}a. Physically it is explained by the increased contribution from the spin torque to the spin
current. When $H\neq 0$, the high-frequency magnetic field flips periodically the electron spins. Since this field is
uniform the difference (which is created by the orbit-orbit coupling) between the group velocities of the electrons moving
in the center of the wire and closer to its edges is not decisive for the ratchet effect. The contribution to the spin
current coming from the group velocity is now smaller than the one due to the spin torque and as a result the orbit-orbit
coupling has a little impact on the spin current.

In summary, we have shown that the intrinsic spatial asymmetry, i.e., the asymmetry not related to the ratchet potential,
of a dissipative system with Rashba spin-orbit interaction can lead to charge and spin ratchet effects when the system is
driven by both electric and magnetic fields. The charge ratchet current has been found to have a purely spin-orbit origin.
The extrinsic spatial asymmetry, i.e., the asymmetry induced by the ratchet potential, is not critical for the existence
of the ratchet effects but its presence amplifies the ratchet currents due to the superposition of the intrinsic and
extrinsic ratchet effects. The proposed system could thus be a multifunctional spintronic device which, when appropriately
electrically programmed by the external periodic gates, works as a spin and/or charge direct current
generator.

\begin{acknowledgments}
Support from the DFG under the program SFB 689 and Excellence Initiative of the German Federal and State Governments is
acknowledged.
\end{acknowledgments}


\begin{thebibliography}{16}
\expandafter\ifx\csname natexlab\endcsname\relax\def\natexlab#1{#1}\fi
\expandafter\ifx\csname bibnamefont\endcsname\relax
  \def\bibnamefont#1{#1}\fi
\expandafter\ifx\csname bibfnamefont\endcsname\relax
  \def\bibfnamefont#1{#1}\fi
\expandafter\ifx\csname citenamefont\endcsname\relax
  \def\citenamefont#1{#1}\fi
\expandafter\ifx\csname url\endcsname\relax
  \def\url#1{\texttt{#1}}\fi
\expandafter\ifx\csname urlprefix\endcsname\relax\def\urlprefix{URL }\fi
\providecommand{\bibinfo}[2]{#2}
\providecommand{\eprint}[2][]{\url{#2}}

\bibitem[{\citenamefont{Astumian and H{\"a}nggi}(2002)}]{Astumian}
\bibinfo{author}{\bibfnamefont{R.~D.} \bibnamefont{Astumian}} \bibnamefont{and}
  \bibinfo{author}{\bibfnamefont{P.}~\bibnamefont{H{\"a}nggi}},
  \bibinfo{journal}{Phys.\ Today} \textbf{\bibinfo{volume}{55}},
  \bibinfo{pages}{33} (\bibinfo{year}{2002}).

\bibitem[{\citenamefont{Reimann et~al.}(1997)\citenamefont{Reimann, Grifoni,
  and H{\"a}nggi}}]{Reimann}
\bibinfo{author}{\bibfnamefont{P.}~\bibnamefont{Reimann}},
  \bibinfo{author}{\bibfnamefont{M.}~\bibnamefont{Grifoni}}, \bibnamefont{and}
  \bibinfo{author}{\bibfnamefont{P.}~\bibnamefont{H{\"a}nggi}},
  \bibinfo{journal}{Phys.\ Rev.\ Lett.} \textbf{\bibinfo{volume}{79}},
  \bibinfo{pages}{10} (\bibinfo{year}{1997}).

\bibitem[{\citenamefont{J{\"u}licher et~al.}(1997)\citenamefont{J{\"u}licher,
  Ajdari, and Prost}}]{Juelicher}
\bibinfo{author}{\bibfnamefont{F.}~\bibnamefont{J{\"u}licher}},
  \bibinfo{author}{\bibfnamefont{A.}~\bibnamefont{Ajdari}}, \bibnamefont{and}
  \bibinfo{author}{\bibfnamefont{J.}~\bibnamefont{Prost}},
  \bibinfo{journal}{Rev.\ Mod.\ Phys.} \textbf{\bibinfo{volume}{69}},
  \bibinfo{pages}{1269} (\bibinfo{year}{1997}).

\bibitem[{\citenamefont{Linke et~al.}(1999)\citenamefont{Linke, Humphrey,
  L{\"o}fgren, Sushkov, Newbury, Taylor, and Omling}}]{Linke}
\bibinfo{author}{\bibfnamefont{H.}~\bibnamefont{Linke}},
  \bibinfo{author}{\bibfnamefont{T.~E.} \bibnamefont{Humphrey}},
  \bibinfo{author}{\bibfnamefont{A.}~\bibnamefont{L{\"o}fgren}},
  \bibinfo{author}{\bibfnamefont{A.~O.} \bibnamefont{Sushkov}},
  \bibinfo{author}{\bibfnamefont{R.}~\bibnamefont{Newbury}},
  \bibinfo{author}{\bibfnamefont{R.~P.} \bibnamefont{Taylor}},
  \bibnamefont{and} \bibinfo{author}{\bibfnamefont{P.}~\bibnamefont{Omling}},
  \bibinfo{journal}{Science} \textbf{\bibinfo{volume}{286}},
  \bibinfo{pages}{2314} (\bibinfo{year}{1999}).

\bibitem[{\citenamefont{Majer et~al.}(2003)\citenamefont{Majer, Peguiron,
  Grifoni, Tusveld, and Mooij}}]{Majer}
\bibinfo{author}{\bibfnamefont{J.~B.} \bibnamefont{Majer}},
  \bibinfo{author}{\bibfnamefont{J.}~\bibnamefont{Peguiron}},
  \bibinfo{author}{\bibfnamefont{M.}~\bibnamefont{Grifoni}},
  \bibinfo{author}{\bibfnamefont{M.}~\bibnamefont{Tusveld}}, \bibnamefont{and}
  \bibinfo{author}{\bibfnamefont{J.~E.} \bibnamefont{Mooij}},
  \bibinfo{journal}{Phys.\ Rev.\ Lett.} \textbf{\bibinfo{volume}{90}},
  \bibinfo{pages}{056802} (\bibinfo{year}{2003}).

\bibitem[{\citenamefont{H{\"a}nggi and Marchesoni}()}]{Haenggi}
\bibinfo{author}{\bibfnamefont{P.}~\bibnamefont{H{\"a}nggi}} \bibnamefont{and}
  \bibinfo{author}{\bibfnamefont{F.}~\bibnamefont{Marchesoni}},
  \eprint{arXiv:0807.1283v1 (accepted in Rev. Mod. Phys.)}.

\bibitem[{\citenamefont{Scheid et~al.}(2007{\natexlab{a}})\citenamefont{Scheid,
  Pfund, Bercioux, and Richter}}]{Scheid}
\bibinfo{author}{\bibfnamefont{M.}~\bibnamefont{Scheid}},
  \bibinfo{author}{\bibfnamefont{A.}~\bibnamefont{Pfund}},
  \bibinfo{author}{\bibfnamefont{D.}~\bibnamefont{Bercioux}}, \bibnamefont{and}
  \bibinfo{author}{\bibfnamefont{K.}~\bibnamefont{Richter}},
  \bibinfo{journal}{Phys.\ Rev.\ B} \textbf{\bibinfo{volume}{76}},
  \bibinfo{pages}{195303} (\bibinfo{year}{2007}{\natexlab{a}}).

\bibitem[{\citenamefont{Scheid et~al.}(2007{\natexlab{b}})\citenamefont{Scheid,
  Bercioux, and Richter}}]{Scheid_1}
\bibinfo{author}{\bibfnamefont{M.}~\bibnamefont{Scheid}},
  \bibinfo{author}{\bibfnamefont{D.}~\bibnamefont{Bercioux}}, \bibnamefont{and}
  \bibinfo{author}{\bibfnamefont{K.}~\bibnamefont{Richter}},
  \bibinfo{journal}{New\ J.\ Phys.} \textbf{\bibinfo{volume}{9}},
  \bibinfo{pages}{401} (\bibinfo{year}{2007}{\natexlab{b}}).

\bibitem[{\citenamefont{Smirnov
  et~al.}(2008{\natexlab{a}})\citenamefont{Smirnov, Bercioux, Grifoni, and
  Richter}}]{Smirnov}
\bibinfo{author}{\bibfnamefont{S.}~\bibnamefont{Smirnov}},
  \bibinfo{author}{\bibfnamefont{D.}~\bibnamefont{Bercioux}},
  \bibinfo{author}{\bibfnamefont{M.}~\bibnamefont{Grifoni}}, \bibnamefont{and}
  \bibinfo{author}{\bibfnamefont{K.}~\bibnamefont{Richter}},
  \bibinfo{journal}{Phys.\ Rev.\ Lett.} \textbf{\bibinfo{volume}{100}},
  \bibinfo{pages}{230601} (\bibinfo{year}{2008}{\natexlab{a}}).

\bibitem[{\citenamefont{Smirnov
  et~al.}(2008{\natexlab{b}})\citenamefont{Smirnov, Bercioux, Grifoni, and
  Richter}}]{Smirnov_1}
\bibinfo{author}{\bibfnamefont{S.}~\bibnamefont{Smirnov}},
  \bibinfo{author}{\bibfnamefont{D.}~\bibnamefont{Bercioux}},
  \bibinfo{author}{\bibfnamefont{M.}~\bibnamefont{Grifoni}}, \bibnamefont{and}
  \bibinfo{author}{\bibfnamefont{K.}~\bibnamefont{Richter}},
  \bibinfo{journal}{Phys.\ Rev.\ B} \textbf{\bibinfo{volume}{78}},
  \bibinfo{pages}{245323} (\bibinfo{year}{2008}{\natexlab{b}}).

\bibitem[{\citenamefont{Rashba}(1960)}]{Rashba}
\bibinfo{author}{\bibfnamefont{E.}~\bibnamefont{Rashba}},
  \bibinfo{journal}{Fiz.\ Tverd.\ Tela (Leningrad)}
  \textbf{\bibinfo{volume}{2}}, \bibinfo{pages}{1224} (\bibinfo{year}{1960}).

\bibitem[{\citenamefont{Dresselhaus}(1955)}]{Dresselhaus}
\bibinfo{author}{\bibfnamefont{G.}~\bibnamefont{Dresselhaus}},
  \bibinfo{journal}{Phys.\ Rev.} \textbf{\bibinfo{volume}{100}},
  \bibinfo{pages}{580} (\bibinfo{year}{1955}).

\bibitem[{\citenamefont{Grifoni et~al.}(2002)\citenamefont{Grifoni, Ferreira,
  Peguiron, and Majer}}]{Grifoni}
\bibinfo{author}{\bibfnamefont{M.}~\bibnamefont{Grifoni}},
  \bibinfo{author}{\bibfnamefont{M.~S.} \bibnamefont{Ferreira}},
  \bibinfo{author}{\bibfnamefont{J.}~\bibnamefont{Peguiron}}, \bibnamefont{and}
  \bibinfo{author}{\bibfnamefont{J.~B.} \bibnamefont{Majer}},
  \bibinfo{journal}{Phys.\ Rev.\ Lett.} \textbf{\bibinfo{volume}{89}},
  \bibinfo{pages}{146801} (\bibinfo{year}{2002}).

\bibitem[{\citenamefont{Goychuk and H{\"a}nggi}(1998)}]{Goychuk}
\bibinfo{author}{\bibfnamefont{I.}~\bibnamefont{Goychuk}} \bibnamefont{and}
  \bibinfo{author}{\bibfnamefont{P.}~\bibnamefont{H{\"a}nggi}},
  \bibinfo{journal}{EPL} \textbf{\bibinfo{volume}{43}}, \bibinfo{pages}{503}
  (\bibinfo{year}{1998}).

\bibitem[{\citenamefont{Caldeira and Leggett}(1981)}]{Caldeira}
\bibinfo{author}{\bibfnamefont{A.~O.} \bibnamefont{Caldeira}} \bibnamefont{and}
  \bibinfo{author}{\bibfnamefont{A.~J.} \bibnamefont{Leggett}},
  \bibinfo{journal}{Phys.\ Rev.\ Lett.} \textbf{\bibinfo{volume}{46}},
  \bibinfo{pages}{211} (\bibinfo{year}{1981}).

\bibitem[{\citenamefont{Weiss}(2008)}]{Weiss}
\bibinfo{author}{\bibfnamefont{U.}~\bibnamefont{Weiss}},
  \emph{\bibinfo{title}{Quantum Dissipative Systems}}
  (\bibinfo{publisher}{World Scientific, Singapore}, \bibinfo{year}{2008}),
  \bibinfo{edition}{3rd} ed.

\end{thebibliography}
\end{document}